                                 \documentstyle[12pt,aaspp4]{article}
\def\gtrsim{\mathrel{\hbox{\rlap{\hbox{\lower4pt\hbox{$\sim$}}}\hbox{$>$}}}}
\let\ga=\gtrsim
\def\lesssim{\mathrel{\hbox{\rlap{\hbox{\lower4pt\hbox{$\sim$}}}\hbox{$<$}}}}
\let\la=\lesssim
                                \begin{document}

                                \title{
Velocity Gradients in the Intracluster Gas of the Perseus Cluster
                                }

                                \author{
Renato A. Dupke \& Joel N. Bregman
                                }

                                 \affil{
University of Michigan, Ann Arbor, MI 48109-1090
                                }

                                \begin{abstract}
We report the results of spatially resolved X-ray spectroscopy of 8 different {\sl ASCA}
pointings distributed symmetrically around the center of the Perseus cluster. The 
outer region of the intracluster gas is roughly isothermal, with temperature $\sim$ 6--7 
keV, and metal abundance $\sim$ 0.3 Solar. Spectral analysis of the central pointing
 is consistent with the presence of a cooling flow and a central metal abundance gradient.
  A significant velocity gradient is found along an axis at a position angle of $\sim$ 135$^\circ$,
  which is $\sim$ 45$^\circ$ discrepant with the major axis of the X-ray elongation. 
The radial velocity difference is found to be greater than 1000 km s$^{-1}Mpc^{-1}$ at the 
90\% confidence level. Simultaneous 
fittings of GIS 2 \& 3 indicate that the velocity gradient is significant at the 95\% confidence level 
and the F-test rules out constant velocities 
at the 99\% level. Intrinsic short and long term variations of 
gain are unlikely (P $<$ 0.03) to explain the velocity discrepancies.

                                \end{abstract}

                                \clearpage
                                \section{
Introduction
                                }

X-ray determination of the physical state of the intracluster gas provides a 
unique tool to probe the origin and evolution of clusters of galaxies.  {\sl ASCA} 
observations have shown significant spatial temperature variations in many clusters, 
indicating that clusters are currently evolving systems. This is consistent with the 
predictions of hierarchical cluster formation within CDM models, and had been 
suggested in pre-{\sl ASCA} times (e.g. Fitchett 1988, Ulmer, Wirth \& Kowalski 1992 
and references therein).

Although early models of galaxy clusters treated them as spherically symmetric virialized systems, 
recent X-ray and optical studies show that often clusters show substructures. 
Furthermore, if cold dark matter models (CDM) are correct, primordial small-scale density 
fluctuations are not erased and clusters are formed by infall/merging of smaller scale 
objects (bottom-up hierarchical scenario). The merger (infall) of sub-clumps creates 
shocks (associated with temperature substructure), bulk gas flows (associated with velocity 
substructure) and asymmetric distributions of velocity of galaxies (e.g. Bird 1993). 
In order to understand the physical properties of clusters, their origin and evolution one has 
to take into account the degree and the physical scale of substructuring. Furthermore, the 
degree of substructuring itself can be used to determine/constrain cosmological parameters 
(Crone, Evrard \& Richstone 1996).

Measurements of substructure using spatially resolved X-ray spectra have some advantages 
over optical analysis of galaxies in clusters. Firstly, one does not need a 
large number of galaxies' redshifts in a cluster. Secondly, the velocity mapping of 
the ICM is not biased by the inclusion of foreground/background galaxies (outliers), 
which may bias the statistical analysis for measuring substructures (e.g. Fitchett 1988, Bird 1993). 
The determination of complex temperature substructure in clusters is often  
interpreted to be related to shocks due to cluster merger at different stages. 
The link between temperature substructure and the merger stage is often done 
by comparison with hydrodynamical simulations. There are currently an enormous 
variety (different initial conditions, hydro-codes, impact parameters, matter 
components, etc.) of cluster formation/merger simulations in the literature (e.g. 
Evrard 1990; Katz \& White 1993; Roettiger, Burns \& Loken 1993,1996; Schindler \& Muller 1993; 
Pearce, Thomas \& Couchman 1994; Navarro, Frenk \& White 1995; Evrard, Metzler, \& Navarro 1996; 
Roettiger, Loken \& Burns 1997, Ricker 1998; Takizawa \& Mineshige 1998; 
Takizawa 1999, 2000 and references therein), which can be used to compare with the observations. 

A more straightforward way to determine the effects of a merger is to directly measure 
 intracluster gas bulk velocities. Several simulations of cluster mergers 
indicate residual gas velocities of a few thousand km s$^{-1}$ (e.g. Ricker 1998, 
Roettiger et al. 1993, Takizawa 1999, Roettiger \& Flores 2000). To measure gas velocities one 
requires accurate determinations of spectral 
line centroids. The precision with which a line centroid can be measured, in velocity space, 
is $\sim 127~\Gamma_{eV}(E_{keV} N^{\frac{1}{2}})^{-1}$ km~s$^{-1}$, where N is the number of photons in 
the line and $\Gamma_{eV}$ is the FWHM of the line, or if the line is narrower than 
the instrumental width, is the FWHM of the instrument, and E$_{keV}$ is the line energy.  
The energy resolution of the spectrometers on-board {\sl ASCA} 
 vary from 2\% (SIS) to 8\% (GIS)
at 5.9 keV. For a FWHM of 9000 km~$s^{-1}$ at 6.7 keV,  which is typical of early (first 3 years) 
SIS data at the FeK$\alpha$ line), to obtain a line centroid to a precision of 500 km~$s^{-1}$ we need 
$\sim$ 60 line photons.  This same accuracy can be obtained with the GIS with $\sim$ 350 line photons
\footnote{These estimates become more uncertain if the rms of the continuum starts to compete with the line.}. 
Although several clusters observed by
{\sl ASCA} match this requirement, when observing regions within the detector field of view, the flux in the 
outer regions of the detector is dominated by photons coming from central regions due 
to the extended {\sl ASCA} PSF and if there are strong radial velocity gradients the latter effect would make
redshift determinations unprecise. Therefore, ideally, one would like to analyze high-metal abundance clusters 
that have several different long-exposure observations from off-center regions surrounding the cluster's X-ray center.
Some nearby clusters match this criteria and, therefore, are well suited for the study of gas velocity distribution. 
Perseus is an ideal target 
for this analysis since it is a nearby very bright cluster. Furthermore, it has been observed extensively by {\sl ASCA} 
with a large number of offset pointings, covering a more or less symmetrical region around the cluster's center
up to distances of $>$ 1 h$_{50}^{-1}$ Mpc. 

In this work we analyze the spectra of 8 separate pointings encompassing a region 
of $>$ 40$\arcmin$ radius around the center of the Perseus cluster. 
We find that although the cluster is roughly isothermal (aside from the cooling flow) there 
is a significant 
velocity gradient along a direction that has an inclination $\ge$ 50$^{\circ}$ with 
respect to the cluster's X-ray major axis. This velocity gradient is consistent 
with a rotation velocity of $\ge$ 1000 km s$^{-1}$  at the 90\% confidence level and is 
unlikely to be caused by gain variations or background sources.

                                \section{
The Perseus Cluster
                                }

The Perseus cluster has been the subject of many studies since its discovery as 
an X-ray source by Fritz et al. (1971). It is one of the closest (at an optical 
redshift of 0.0183), X-ray bright, rich cluster of galaxies. The cluster is 
elongated and the ratio of its minor to major axis is 0.83 at radii greater than 
20$\arcmin$ (Snyder et al. 1990). It has a cooling flow with a mass deposition rate 
of about (3--5) $\times 10^{2}$ M$_{\odot}$ yr$^{-1}$(Allen et al. 1992; Peres et al. 
1998; Ettori, Fabian \& White 1998).  The 
centroid of the cluster emission is offset by $\sim$ 2$\arcmin$ to the east of NGC 1275 
(Snyder et al. 1990, Branduardi-Raymond et al. 1981). The average temperature of 
the X-ray emitting gas is approximately 6.5 keV (Eyles et al. 1991) and the average 
abundance is 0.27 Solar in the central 1 degree (Arnaud et al. 1994). 

The existence of an iron abundance gradient in Perseus was first suggested by 
Ulmer et al. (1987) using data from SPARTAN 1. They found an iron abundance of $\sim$ 0.81
 Solar and a temperature of $\sim$ 4.16 keV within the central 5 $\arcmin$ and an abundance 
 of $\sim$ 0.41 solar and a temperature of $\sim$ 7.1 keV in the outer regions 
(6 - 20$\arcmin$). Further analyses have corroborated the existence of an abundance gradient
(e.g. Ponman et al. (1990), Kowalski et al. (1993), Arnaud et al. 1994, 
Dupke \& Arnaud 2000). Furthermore, the region where the abundance is enhanced is 
predominantly enriched by SN Ia ejecta (Dupke \& Arnaud 2000), indicating that the 
cluster belongs to the class of clusters that present central ``chemical gradients'', 
such as A496 (Dupke \& White 2000). The presence of cooling flows, global abundance and chemical
gradients suggest that the cluster has not undergone strong mergers recently.

                                \section{
Data Reduction \& Analysis
                                }

{\sl ASCA} carries four large-area X-ray telescopes, each with its own detector: two Gas 
Imaging Spectrometers (GIS) and two Solid-State Imaging Spectrometers (SIS). Each 
GIS has a $50'$ diameter circular field of view and a usable energy range of 0.8--10 
keV; each SIS has a $22'$ square field of view and a usable energy range of 0.5--10 keV.
The nominal energy resolution of the spectrometers are 8\% and 2\% at 5.9 keV for GISs 
and SISs, respectively. The SISs energy resolution is steadily degrading with time 
(e.g. Dottani et al. 1997) and for most pointings 
analyzed in this work is $\ge$  4\% at 5.9 keV. Eight individual pointings were analyzed 
in this work. The central pointing is the only one that includes the contaminating 
source (the center of the Perseus cluster). The other seven pointings are distributed
more or less symmetrically around the central pointing with an average distance of 
$\sim$ 40$\arcmin$ from the center. Five of the pointings (the most recent ones) were 
consecutively observed in 1997 and are spaced in time typically by a day. The pointing
characteristics are listed in Table 1 and shown in Figure 1. 

We selected data taken with high and medium bit rates, with cosmic ray rigidity
values $\ge$ 6 GeV/c, with elevation angles from the bright Earth of $\ge20^{\circ}$, 
and from the Earth's limb of $\ge5^{\circ}$ (GIS); we also excluded times when the 
satellite was affected by the South Atlantic Anomaly. Rise time rejection of particle 
events was performed on GIS data. The resulting effective exposure times are also shown in 
Table 1. We estimated the background from blank sky files provided by the {\sl ASCA} 
Guest Observer Facility and removed bright point  sources for each instrument in all 
pointings.

The SISs have a better spectral resolution (by a factor of 2--4) than the 
GISs. However, the analysis of SIS data for our pointings is severely limited by 
photon statistics (the GISs overall count rate is typically more than ten times that of the SISs),
making the SIS redshift determination very uncertain. This difference in count rate between 
GISs and SISs is due to the following reasons: 1) most of the pointings analyzed in 
this work were observed by the SIS in 2-CCD mode, which minimizes the energy resolution 
degradation due to the residual dark-current distribution and non-uniform charge 
transfer inefficiency effects\footnote{heasarc.gsfc.nasa.gov/docs/asca/newsletters/sis\_calibration5.html}; 
2) there  have been 
increasing discrepancies between SIS and GIS spectra below 1 keV since 1994, due to 
a decrease in X-ray efficiency in the SISs, thus making the low end ($<$0.9 keV)of the spectral 
region unusable for the SISs\footnote{heasarc.gsfc.nasa.gov/docs/asca/watchout.html, also Hwang et al. 1999};
3) the X-ray center of Perseus is not in the detector's field of view for all outer pointings
analyzed and, therefore, most of the photons come from a 
specific direction (towards the cluster's center).  Furthermore, the GISs have a higher effective area at high 
energies. Therefore,  we use only GIS 2 \& 3 in this analysis. 

In the spectral fittings we used {\tt XSPEC} v10.0 (Arnaud 1996) software to analyze the GIS spectra 
separately and jointly (simultaneous fittings of data from GIS 2\&3). The spectra were fitted 
using the {\tt mekal} thermal emission
models, which are based on the emissivity calculations of  Mewe \& Kaastra (cf. Mewe,
Gronenschild \& van den Oord  1985; Mewe, Lemen \& van den Oord 1986; Kaastra  1992),
with Fe L calculations by Liedahl, Osterheld \& Goldstein (1995). Abundances are 
measured relative to the solar photospheric values of Anders \& Grevesse (1989), 
in which Fe/H=$4.68\times10^{-5}$ by number. Galactic photoelectric absorption was 
incorporated using the {\tt wabs} model (Morrison \& McCammon 1983); Spectral channels 
were grouped to have at least 25 counts/channel. Energy ranges were restricted to 
0.8--9 keV for the GISs. 

Since there is a cooling flow at the center of Perseus we added a cooling 
flow component (Mushotzky \& Szymkowiak 1988) to the {\tt mekal} thermal emission model in the central pointing, 
to compare the temperature of the hot component in the center with that of the outer
pointings. We adopted the emission measure temperature distribution that corresponds 
to isobaric cooling flows. We tied the maximum temperature of the cooling flow to the 
temperature of the thermal component, and we fixed the minimum temperature at 0.1 keV. 
The abundances of the two spectral components ({\tt mekal} and {\tt cflow}) were tied
together. We also applied a single (but variable) global absorption to both spectral 
components. Since the cD galaxy of Abell 426 (NGC 1275) is an AGN, we also included a power
law component in the spectral fittings of the central pointing.

We extracted spectra from a circular region with a radius of 20$\arcmin$ for each pointing, centered
at the detector's center. The best-fit values for temperature, abundance and redshift obtained 
from spectral 
fittings of GIS2 and GIS3 separately are consistent with those obtained through the 
spectral fittings of both GIS 2 \& 3 jointly. Therefore, we show here only the best-fit 
parameters of the simultaneous fittings. The resulting joint fits were very good with 
reduced $\chi^2_{\nu}\sim 1$ for all regions. 

                                        \section{
Results
                                         }

                                        \subsection{
Spectral Fittings
                                        }

The best-fit values for temperature, abundance and redshift are shown in Table 2, 
and plotted in Figure 2 as a function of the azimuthal angle. Here, we define the azimuthal angle with
respect to the line that joins the centers of pointings P1(NW) and P5(SE), for convenience. For P1(NW) the azimuthal angle is set to zero.
 The indicated errors in Figure 2 are 
$1\sigma$ confidence limits. It can be seen that Perseus appears to be roughly isothermal in
the outer regions  with an average temperature of 
$\sim$ 7 keV. The dashed and solid lines for the temperature plot in Figure 2 
indicate the $1\sigma$ confidence levels for the simultaneous GIS 2 \& 3 spectral fittings 
of the central pointing with and without the absorbed cooling flow component, 
respectively. The central best-fitting temperature for models that include a cooling 
flow component is not well constrained by the GISs and shows a best-fit value of 
6.85$\pm1$ keV, which is consistent with the observed temperatures in the outer 
regions. Since the GISs are not very sensitive to the absorbing column density the 
absolute values of the best-fitting temperatures may be artificially increased if 
the best-fit N$_{H}$ values are low.
 To test this effect we fixed N$_{H}$ at its putative Galactic value at the 
direction of each pointing (N$_{H}~\sim 1.6\times 10^{21} cm^{-2}$, Dickey \& Lockman 1990;
HEASARC NH Software). 
The best-fitting temperatures obtained this way have  significantly worse $\chi^{2}_{\nu}$ 
and are also 
plotted in Figure 2 (open circles). The best-fit temperatures obtained when 
N$_{H}$ is free to vary are lower than those obtained with free N$_{H}$ by $\sim$ 1.5 keV. 
Although the azimuthal 
distribution of temperatures is consistent with isothermality, some 
pointings show mild, but significant, azimuthal variations (e.g. Pointing 7(W), 
for which the temperature 
is significantly ($>$90\% confidence) higher than P6 (S-SW) and P3 (NE) by about 
a keV).

The abundances observed in the outer parts of Perseus are generally lower than 
in the central region, which is consistent with observations by other authors (Ulmer et al. 1987; 
Ponman et al. 1990; Kowalski et al. 1993; Arnaud et al. 1994; Dupke \& Arnaud 2000). The 
abundance measured 
in the central pointing is 0.43$\pm$0.02 Solar and in the outer parts has an average value of 
$\sim$0.33 Solar. There is marginal evidence of azimuthal abundance variations. In particular,  
P6 (S-SW) best-fit abundance is higher than the  
that measured for P2 (N), and is also consistent with the abundance measured 
in the central region. The best-fit abundances for each pointing are also displayed 
in Figure 2, where the solid lines represent the $1\sigma$ confidence limits for the 
abundance in the central pointing. The dash-dotted lines show the 90\% limits for 
the abundance measured within a 4$\arcmin$ region of the center of Perseus for comparison 
(Dupke \& Arnaud 2000).

The most important azimuthal distribution is that of redshifts. Two pointings 
show significant ($\ge$ 90\% confidence level) discrepant redshifts with respect to the 
best-fit redshift observed in the central pointing. These two redshift-discrepant 
pointings, P1(NW) and P5(SE), are on opposite sides of the cluster's center. 
P1(NW) shows a best-fit redshift of 0.0145 (0.003-0.0185) and P5(SE) shows a significantly 
higher (95\% confidence level) redshift value of 0.042 (0.025-0.045). This redshift discrepancy is observed in both
GIS 2 and GIS 3 individually, although with lower statistical significance. 
This differences in redshifts imply a velocity difference of 8200 (2000-12700) 
km s$^{-1}$ between these two pointings. The azimuthal distribution of redshifts is shown 
in Figure 2 and the best-fit values are listed in Table 2. In the 
fittings where the column density is fixed at the Galactic value, the best-fit 
redshifts are typically lower than those obtained with N$_{H}$ free. However, the 
inferred differences between redshifts for different pointings are virtually unaltered. 
Therefore, the redshift discrepancies are not due to uncertainties related to the GIS
sensitivity to hydrogen column densities. 
 
The two redshift-discrepant pointings show no differences in the best-fit values of 
temperatures or abundances. To further test the significance of the velocity 
difference between P1 and P5, we simultaneously fit all four spectra (GIS 2 \& 3 for
each of the two pointings) and applied the F-test in the analysis of $\chi^2$ variations due to 
the change in the number of degrees of freedom. We compare the $\chi^2$ of fits which 
assumed the redshifts were the same in the two projected spatial regions 
$z_{G2_{P1}} = z_{G3_{P1}} = z_{G2_{P5}} = z_{G3_{P5}}$
to that of fits which allowed the redshifts in the two pointings to vary independently. 
Within the same pointing the redshifts were still locked together for different 
instruments, i.e., $(z_{G2_{P1}} = z_{G3_{P1}}) \ne (z_{G2_{P5}} = z_{G3_{P5}})$, where $z_{Gi_{Pj}}$ is
the redshift of pointing $j$ with instrument GIS $i$.
The difference between the $\chi^2$ of these two fits must follow a $\chi^2$ 
distribution with one degree of freedom (Bevington 1969). 
The difference in $\chi^2$ from the fits with locked and unlocked redshifts indicates 
that the redshifts are discrepant at the $99$\% level. We show in Figure 3 the 68\%, 90\% and 99\% 
confidence contours for two interesting parameters (redshifts) of regions P1 and P5 as well as
the line correspondent to equal redshifts.
The spectral fits within the energy range encompassing 
the FeK line complex are shown in Figures 4a,b. We also show in Figures 4a,b the residuals 
from the same model fittings but with zero metal abundances for illustration.

The inclusion of the power law component in the spectral fittings (representing any 
non-thermal emission from NGC 1275) in addition to the cooling flow component in
the central region (P0) does not change significantly the best-fit values of the redshifts 
obtained without the power law component. However, it does makes the best-fit redshift values 
less precise. Since the F-test shows that there is a significant improvement 
in the spectral fittings when the power-law component is included 
we, conservatively, quote in Table 2 and Figure 2
the values of the best-fit redshift for spectral fittings that included an extra power 
law component.
                                \subsection{
GIS Gain Variations
                                }

We have shown in the previous section that the azimuthal distribution of velocities
in the Perseus cluster shows significantly ($\ge$ 90\% confidence) different redshifts 
for three Pointings: the central pointing (P0) and two other diametrically opposed pointings
(P1 \& P5).
However, since the determination of redshifts relies on accurate measurement of line centroids 
(mainly the FeK line around 6.7 keV), large variations of gain (conversion between photon 
energy and pulse height) can, in principle, mimic the observed effect. In this section, we 
estimate the effects of gain variations in our observations.

The original 
gain calibration of the GISs was mainly based on the built-in Fe-55 isotope source, 
attached to the edge of the field of view. The gain depends on the temperature of 
the phototube ($\sim$1\%/$^{\circ}$C), the position on the detector and time of 
observation. During the first several months in orbit the gain decreased by a few 
percent, and this trend has slowly disappeared. This gain decrease is possibly 
due of cosmic-ray induced crystalline defects, decreasing 
the UV transmission of the quartz windows of the gas cell (Tashiro et al. 1995). 

The intrinsic GIS gain is not only dependent on temperature but also on position 
(on the detector) due to non-uniformity in the phototube gain. Therefore, the gain 
correction process involves a look-up table called the `gain map' 
(Tashiro et al. 1995 and references therein),
which is also dependent on time and has been recalibrated using spectral lines 
observed during long ``day Earth'' and ``night Earth'' observations.
This allowed for more precise measurement of the azimuthal variation of gain across 
the detector, which is typically smaller than the radial gain variations.

The four gain corrections (short and long term gain variation, gain positional 
dependence and long term variation of the positional dependence), were carried 
out at GSFC in the standard processing (Ebisawa, private communication). 
The Perseus observations analyzed in 
this work do have the standard gain corrections applied with the ftool {\tt ASCALIN} v0.9t, 
which reads the gain correction coefficients at the time of observation from the gain 
history file.  The gain correction coefficients were created by the ftool 
{\tt TEMP2GAIN} v4.2, which reads the variation of the temperature housekeeping parameters, 
and registers the gain values for every 600 seconds of observation taking into account 
the long-term and positional gain variations (Idesawa et al. 1995).

Since there are still observed small redshift fluctuations measured with GIS 2 and, especially, GIS 3 for the same 
region, we assume that, in spite of the standard gain correction, there are 
still residual gain variations and we also assume, conservatively, that the 
magnitude of the residual gain fluctuations are on the same order as the fluctuations of 
the gain observed using the instrumental copper fluorescent line at 8.048 keV 
(Tashiro et al. 1999). Since the gain fluctuations increase 
with time, we use the 1997 data as our standard of reference. The radial gain 
distribution shows that if one excludes the very outer region 
(r $\ge$ 22$\arcmin$) and the very central region ($\le$ 2.5$\arcmin$) from the spectral 
analysis the gain fluctuation around the mean is $\le$ 0.15\%, for both GIS 2 \& 3. For all pointings 
observed in this work we extract spectra from a circular region with a 20$\arcmin$ 
radius. The exclusion of the central 2.5$\arcmin$ from the pointings with discrepant redshifts 
doesn't change our results. This is not surprising, since they are offset pointings 
(most photons come from a specific direction and not from the detector's center). 

However, the direction from which most photons are detected for pointings P1 and P5 are 
different, so we also need to estimate the azimuthal gain 
variations. We also used the gain map determined using the Cu-K line at 8.048 keV 
(Idesawa et al. 1995). We compare the average gain values (excluding 
the outermost ring) of the region encompassing a 90$^{\circ}$ slice corresponding to 
the direction towards the real cluster's center (which is out of the field of 
view) for each instrument. This should give a good idea about the magnitude of the gain 
fluctuations as a function of detector's position for our observations. Although the gain 
differences for the GIS 2 obtained this way imply a small gain variation ($\sim$ 0.12\%), for the 
GIS 3 the derived gain fluctuation is substantially larger ($\sim$ 0.37\%) than
that observed for the radial variations.

                           \subsection{
Redshift Dependence on Gain
                                }

In order to test the sensitivity of our observations to possible residual gain 
variations across the GISs we used Monte Carlo simulations. Supposing the redshift 
to be constant for the two discrepant regions, we generated fake spectra for both 
the GIS 2 \& 3 for the two pointings and compare the best-fit redshift differences.
We can then calculate the probability that we find the same redshift differences 
(or greater than) that we observe in the real pointings for GIS 2 \& 3.  We simulated 
1000 GIS 2 \& 3 spectra corresponding to each real observation using {\tt XSPEC} tool {\tt fakeit}. 
In generating the fake spectra we used the same spectral model ({\tt wabs} {\tt mekal}) that 
we used for fitting the real data with the values of temperature, abundance, N$_{H}$ 
and normalizations correspondent to the real best-fitting values of each pointing. 
We used the actual background spectra extracted within the same spatial extraction
region for each pointing and effective exposures corresponding to the real pointing 
being simulated. We also used the responses (ARF and RMF) corresponding to the real
pointings. Count statistics were incorporated in the generated spectra. For all 
simulations we set the redshift to that obtained through the spectral fittings of 
the central pointing (z = 0.025).

Each simulated spectrum was then grouped (25 cnt/channel) and fitted in the same way as the real 
ones and the best-fit values of the redshift were recorded. We then selected the 
simulated spectra that had a redshift difference equal to or greater than 
that observed in the real pointings for both GIS 2 \& 3 (0.0208 for GIS2 and 
0.0302 for GIS3). Gain effects do not enter directly in the procedure of 
generating simulated spectra. Therefore, in order to estimate the effects 
of GIS gain variations, we added a gain uncertainty to the best-fit 
values of redshift derived from fake spectra. We assume that the gain variations 
follow a Gaussian distribution with a standard deviation ($\sigma_{gain}$), 
which is different for each spectrometer, and a zero mean. This gain uncertainty 
is then summed to the best-fit redshifts obtained from the fake spectra before calculating the 
redshift differences between different pointings. To be conservative we assumed as 
our 1-$\sigma$ gain variations ($\sigma_{gain}$) for the GIS 2 \& 3 
the largest values of the two procedures described above, i.e. 0.15\% and 0.37\% for 
GIS 2 \& GIS 3, respectively.

The probability of observing the redshift differences in GIS 2 \& 3 that we measure 
for the real spectra in two pointings (P1 and P5), using the procedure described 
above is found to be $\la$ 0.005. To illustrate how sensitive this value is to the assumed 
$\sigma_{gain}$ we varied the estimated $\sigma_{gain}$ and recalculated the 
probability of finding the redshift differences by chance. We plot the results 
in Figure 5 (for the purpose of illustration $\sigma_{gain}$ for GIS 2 \& 3 are 
assumed to be the same). It can be seen that this probability is rather insensitive 
to gain variations up to a $\sigma_{gain}$ of $\sim$ 0.5\%.

In a more realistic case we observe overall seven outer pointings and not only two. Therefore, 
we estimated the probability of finding the same redshift differences that we see 
in pointings P1 and P5 in seven observations using the same procedure to simulate 
spectra as described above. We also included a condition for alignment. Consider the 
line that crosses a pointing Pj (which has a azimuthal angle PAj) and the X-ray center of Perseus. 
Any pointing Pi (with azimuthal angle PAi) will 
be considered to be aligned with pointing Pj if (PAj+$\pi$-A) $\le$ PAi $\le$ (PAj+$\pi$+A),
where A is some alignment angle. Conservatively, we assume a broad alignment condition 
(A=$ \frac {\pi}{3}$). 
The results are also shown in Figure 5 (thin line). Even in this case, the redshift
difference  is still significant at $\ga$~97\% confidence level.

                           \section{
Discussion
                                }

The spectral analysis of {\sl ASCA} GIS 2 \& 3 carried out in this work indicates the 
presence of a significant velocity gradient in the ICM of the Perseus cluster. Two regions show 
discrepant redshifts not just with respect to each other but also with respect to the central region.
We have shown in the previous paragraphs that this difference is unlikely to be 
attributed purely to gain fluctuations and suggests the existence of large-scale bulk motions of 
the intracluster gas in this cluster. The temperature measurements show a general
radial positive gradient, consistent with a cooling flow in the inner regions and an 
isothermal distribution in the outer regions. The abundance distributions are 
consistent with a global central abundance enhancement decreasing by about 30\% 
outwards. 

The two symmetrically opposed discrepant regions have velocity differences of $\sim$~-3000 
km~s$^{-1}$ (or $\le$ -~600 km~s$^{-1}$ at the 90\% confidence level) for P1 and 
$\sim$ +5000 km~s$^{-1}$ (or $\ge$ +~60 km~s$^{-1}$ at the 90\% confidence level) 
for P5 with respect to P0. There are no observed temperature or abundance differences 
in the two pointings with discrepant redshifts (P1 \& P5).
The velocities measured are consistent with large scale gas rotation with a correspondent
circular velocity of $\sim$ 4100 $^{+2200}_{-3100}$ km s$^{-1}$ (90\% confidence). This  
implies a large angular momentum for the ICM
and that a significant fraction of the gas energy is kinetic\footnote{
Implying a correction on the measured specific energy of the 
gas, lowering $\beta_{spec}$ and the $\frac{\beta_{spec}}{\beta_{imag}}$ 
discrepancy (Evrard 1990; Allen et al. 1992).}. 

The best candidate for generating this large angular momentum is off-center 
mergers. In off-center mergers, up to $\sim$ 30\% of the total merger energy may be kinetic
(can be transferred to rotation) (e.g. Pearce et al. 1994). Off-center merger simulations 
often produce residual intracluster gas rotation with velocities 
of a few $\times 10^{3} $km$~s^{-1}$ (e.g. Ricker 1998, Takizawa 1999, 2000, Roettiger \& Flores 2000).
Additional evidence for merging in Perseus comes from the observed: 1) offset between the optical 
center and the X-ray center (Branduardi-Raymont et al. 1981; Snyder et al. 1990; Ulmer et al. 1992), 
2) radial change in X-ray isophotal orientation (isophotal twist)(Mohr, Fabricant, \& Geller 1993) and 
3) asymmetric galaxy morphological distribution, with preferential eastward direction in the 
distribution of E+S0 types (Brunzendorf \&  Meusinger 1999). 
However, simulations also indicate other observable consequences of mergers that can, in 
principle, be cross-analyzed with the velocity maps to test the robustness of the merger 
scenario. One of the features predicted by off-center cluster-cluster mergers is a 
strong negative radial temperature gradient (core heating) ($\ge$ 2 keV/Mpc) 
for most of the merger life-time, even when the angle of view is not favored, e.g. along the collision 
axis (e.g. Takizawa 2000, Ricker 1998). 
In our case we do not observe a negative temperature gradient at all. Actually, we 
observe a positive temperature gradient due to the cooling flow. The mere fact that 
the cooling flow is present makes the off-center merger explanation more uncertain, since it 
has been suggested that a merger would disrupt any pre-existing cooling flows 
(e.g. Edge, Steward \& Fabian 1992, Roettiger et al. 1993). However,
recent simulations of head-on cluster mergers indicate that cooling flows can survive
mergers depending on the produced ram-pressure of the gas in the infalling cluster (Gomez et al. 2000).
Even if cooling flows are disrupted in a merger they can be reestablished quickly if the cooling time
of the primary pre-merger component is small (Gomez et al. 2000). The fact that the major 
axis of the X-ray elongation is relatively close to the apparent 
rotation axis is another difficulty posed to the off-merger explanation. In most cases the 
isodensity contours are elongated perpendicularly to rotation axis, except in some 
short-lived merger stages viewed from specific directions (e.g. Takizawa 2000). 

Two options to try to conciliate the velocity gradients that we find with merger scenarios 
are: 1) we are seeing a pre-merger stage with an infalling sub-group; 
2) The merger happened long ago and the cluster was able to reestablish a cooling flow 
(and create a central metal abundance gradient) or the central region has not been disrupted. 
Against the former scenario is that we do not observe X-ray surface brightness 
enhancement at the direction of the redshift-discrepant  regions. {\sl ROSAT} PSPC images of 
P5 do not show any source bright enough to contaminate the overall flux and, consequently, 
the measured redshifts. However, we do observe a mild enhancement in surface brightness 
towards the East region of Perseus, coinciding with our P4 pointing. This enhancement was noticed 
previously by Schwarz et al. (1992) and  Ettori et al. (1998) with better significance, 
and has been interpreted as evidence for merging group, which could also explain the 
lower temperatures detected towards 
the East by Schwarz et al. (1992). We do not detect a temperature gradient towards 
the East, but since our eastern 
pointing (P4) is in a region of higher N$_{H}$ the lack of sensitivity of GIS to the hydrogen column density 
could mask a small temperature decline. However, in order to match the low temperatures found
by {\sl ROSAT} in P4 N$_{H}$ would have to be significantly higher \footnote{Even when N$_{H}$ was fixed 
to be twice as high for P4 as its correspondent Galactic value, the best-fit temperature found 
in that region was still 4.23$\pm 0.13$ with $\chi_{\nu}^{2} \sim 1.46$}. Although there is 
no clear optical evidence of a sub-group towards the direction of P5 that corroborates 
this scenario, the region of surface brightness enhancement (P4) is associated with a high 
fraction of early-type galaxies (Brunzendorf \& Meusinger 1999). 
Brunzendorf and Meusinger (1999) optically detected a possible cluster at z $\sim$ 0.05 $\sim$ 90$\arcmin$
north of NGC 1275. The position of this ``cluster'' coincides with a surface brightness enhanced region
detected with ROSAT (Kowalski 1994)\footnote{There is also a region of enhanced surface brightness $\sim$ 
100$\arcmin$ to the northeast of NGC 1275, also out of the region covered by {\sl ASCA} 
(Kowalski, Personal Communication)}. However, its location is far north  (beyond the region analyzed by 
{\sl ASCA} and closer to the region where we detect low redshifts, and it is unlikely to 
explain the redshift discrepancy that we find in P5 \& P1. 

The second scenario would imply that large rotational
velocities can be maintained for longer periods of time (comparable to a Hubble time if cooling
flows are disrupted in off-center mergers) 
\footnote{The cooling time in the center of Perseus is $\sim$ 1 Gyr and it is $\sim$ 
10 Gyr at $\sim$ 200 kpc, where a mass deposition rate of $\sim$ 500 M$_{\odot}$yr$^{-1}$
is inferred (Ettori et al. 1998).}.
In some off-center merger simulations high rotational velocities can be maintained for $\ge$ 3 
crossing times (e.g. Ricker 1998). However, typically, the gas velocities are higher towards 
the central cluster's regions, which could not be detected in Perseus. If cooling flows are 
not disrupted in off-center mergers, then the results are consistent with a large 
off-center merger event that took place roughly $\ge$ 4 Gyr (assuming a cluster mass of 
5$\times$10$^{14}$ M$_{\odot}$ at a radius of 1 Mpc (Ettori et al. 1998) and 
that the rotating gas at $\sim$ 7 keV is gravitationally bound). 

Since we cannot measure the X-ray elongation along the line-of-sight to make better 
comparison with simulations and it is not 
clear whether cooling flows and central abundance gradients can actually survive mergers 
(e.g. McGlynn \& Fabian 1984; Fabian \& Daines 1991; Allen et al. 1992; Markevitch et al. 2000; Gomez et al. 2000), and,
 given the necessary poor spatial scale required to
measure gas velocities reliably with the GISs, the results shown in this paper cannot constraint 
different merger scenarios accurately. However, it 
provides new challenges to numerical simulations of cluster mergers. Any merger explanation 
for the velocity differences observed in
Perseus will have to take into account the presence of: 1) a moderately high cooling flow,
2)a central metal abundance gradient 
and 3) a chemical gradient, i.e., a central dominance of SN Ia ejecta (Dupke \& Arnaud 2000). 
Velocity measurements of the intracluster gas with 
{\sl Chandra} and, especially, {\sl XMM-Newton}
satellites will be able determine ICM velocities in the central regions more precisely, thus 
providing information on the gas velocity curve, which will strongly constraint 
cluster-cluster merger models, or suggest alternatives for generating the large 
angular momentum observed.

\acknowledgments We would like to thank, J. Arabadjis, G. Bernstein, M. Sulkanen, M. Ulmer 
 and M. Takizawa for helpful discussions, 
J. Irwin for reading the manuscript and helpful suggestions and especially 
P. Fischer for the many helpful discussions and suggestions. The authors would particularly like 
to thank Dr. K. Ebisawa for providing information about {\sl ASCA} GIS gain calibrations that
was crucial to this work. We acknowledge support 
from NASA Grant NAG 5-3247. This research made use of the HEASARC {\sl ASCA} database and NED.

                                
                                \clearpage

                                \begin{figure}
                                \title{
Figure Captions
                                }

                                \caption{
Distribution of the spatial regions for different pointings analyzed in this work. PSPC surface 
brightness contours of Perseus are overlayed on the central pointing (P0).
The radius of each circular region is 20$\arcmin$.
                                }
                                \caption{
Azimuthal distribution of Temperature (TOP), Metal Abundance(MIDDLE), and Redshift (BOTTOM) as 
a function of the azimuthal angle (0--2$\pi$). First data point from the left for all plots corresponds
to P1, increasing to P7 (last). In the temperature plot solid and dashed lines represent the 
$1\sigma$ confidence limits
for the central pointing (P0) without and with an extra cooling flow spectral component, 
respectively. Circles represent best-fit temperatures when N$_{H}$ is fixed at the Galactic 
value for each pointing. The dotted lines show the $1\sigma$ confidence limits
for the central pointing (P0) without the cooling flow spectral component and N$_{H}$ is fixed at the Galactic 
value. In both abundance and redshift plots solid lines show the 
the $1\sigma$ confidence limits for the central pointing. The dash-dotted lines show 
the best-fit abundance values for the central 4$\arcmin$ obtained by Dupke \& Arnaud 2000, for
comparison. Errors for all plots are $1\sigma$ confidence.
                                }
\caption{
Confidence contour plot for the redshifts measured for pointings P1 \& P5. The three contours 
correspond to 68\%, 90\% and 99\% confidence levels (outwards). The line of equal redshifts 
is also indicated. The contours are found for simultaneous spectral fittings of four data groups 
(P1 GIS 2\&3 and P5 GIS 2\&3), and the redshifts of both instruments are locked together for the same
pointing.
                                }
                                \caption{
Spectral fittings for the region around the FeK line complex for P1 and P5 in both 
GIS 2 (a) \& 3 (b). The bottom plots show the residual to the best-fit model when the
 metal abundance is set to zero for GIS 2(a) \& 3(b).
Data points and best-fit models for P1 are represented by dark lines and for P5 by brighter lines.
                                }
                                \caption{
Probability of finding redshift differences equal or greater than what we 
observe for the real data (P1$\times$P5) by chance as a function of the standard deviation of the gain 
variation. The solid line represents this probability for two pointings. The dotted line 
represents this probability for two out of seven pointings with an alignment angle of 60$^{\circ}$. 
An aligment angle of 90$^{\circ}$ would mean that 2 pointings would be aligned just by been on oposite 
relative hemispheres. $\sigma_{gain}$ for GIS 2 \& 3
are assumed to be equal in this plot for ilustration purposes.
                                }
                                \end{figure}
\clearpage

\begin{deluxetable}{lccccccc}
\small
\tablewidth{0pt}
\tablecaption{Analyzed Pointings}
\tablehead{
\colhead{Pointing} &
\colhead{Sequence}  &
\colhead{Date}  &
\colhead{RA} &
\colhead{DEC} &
\colhead{EXP\tablenotemark{a}} &
\colhead{GIS CNT\tablenotemark{b}}  \nl
\colhead{} &
\colhead{Number} &
\colhead{Observed}  &
\colhead{(2000)} &
\colhead{(2000)} &
\colhead{(ksec)} &
\colhead{(kcount)}
}
\startdata
P1(NW)   & $85002000$ & 1997-02-15 & 03h17m42.07s & +41$^{\circ}$56$\arcmin$54.6$\arcsec$ & 16.6 & 25 \nl
P2(N)   & $85001000$ & 1997-02-14 & 03h19m34.25s & +42$^{\circ}$07$\arcmin$33.2$\arcsec$ & 15.4 & 19 \nl
P3(NE)   & $85000000$ & 1997-02-14 & 03h22m03.53s & +41$^{\circ}$59$\arcmin$05.6$\arcsec$ & 19.7 & 23 \nl
P4(E)   & $83052000$ & 1995-08-19 & 03h22m42.24s & +41$^{\circ}$30$\arcmin$47.9$\arcsec$ & 17.5 & 33 \nl
P5(SE)   & $85003000$ & 1997-02-16 &03h21m47.21s & +41$^{\circ}$02$\arcmin$44.9$\arcsec$ & 14.1 & 20 \nl
P6(S-SW)   & $85004000$ & 1997-02-17 & 03h18m52.46s & +40$^{\circ}$57$\arcmin$10.4$\arcsec$ & 23.4 & 33 \nl
P7(W)   & $85009000$ & 1993-09-15 & 03h17m05.04s & +41$^{\circ}$31$\arcmin$22.4$\arcsec$ & 17.4 & 33 \nl
P0(Center) & $80007000$ & 1993-08-06 & 03h20m08.57s & +41$^{\circ}$36$\arcmin$24.1$\arcsec$ & 11.8 & 183 \nl

\enddata
\tablenotetext{a}{Effective Exposure (Average for GIS 2 \& 3)}
\tablenotetext{b}{Effective Counts (Average for GIS 2 \& 3)}
\end{deluxetable}
\clearpage
\begin{deluxetable}{lccccc}
\small
\tablewidth{0pt}
\tablecaption{Spectral Fittings\tablenotemark{a}}
\tablehead{
\colhead{Pointing} &
\colhead{Temp.}  &
\colhead{Temp.\tablenotemark{b}}  &
\colhead{Abundance} &
\colhead{Redshift} &
\colhead{$\chi_{\nu}^{2}$}  \nl
\colhead{} &
\colhead{(keV)} &
\colhead{(keV)}  &
\colhead{(Solar)} &
\colhead{(10$^{-2}$)} &
\colhead{}
}
\startdata
P1(NW)   & 7.28$^{+0.44}_{-0.53}$& 5.41$^{+0.22}_{-0.15}$ & 0.32$^{+0.07}_{-0.07}$& 1.45$^{+0.40}_{-1.16}$ & 1.054\nl
P2(N)   & 6.96$^{+0.48}_{-0.47}$ & 5.19$^{+0.21}_{-0.19}$ & 0.23$^{+0.08}_{-0.07}$ & 1.81$^{+1.20}_{-1.06}$& 1.097 \nl
P3(NE)   & 6.58$^{+0.49}_{-0.48}$ & 4.99$^{+0.22}_{-0.19}$ & 0.34$^{+0.07}_{-0.08}$ & 1.6$^{+1.23}_{-0.87}$ & 1.115\nl
P4(E)   & 6.89$^{+0.41}_{-0.35}$ & 5.46$^{+0.15}_{-0.15}$ & 0.38$^{+0.05}_{-0.06}$ & 2.64$^{+0.46}_{-0.76}$ & 0.988 \nl
P5(SE)   & 7.24$^{+0.56}_{-0.48}$ & 5.78$^{+0.27}_{-0.23}$ & 0.32$^{+0.07}_{-0.07}$ & 4.19$^{+0.33}_{-1.67}$ & 1.006 \nl
P6(S-SW)   & 6.65$^{+0.39}_{-0.33}$ & 5.47$^{+0.16}_{-0.12}$ & 0.40$^{+0.06}_{-0.06}$ & 2.19$^{+0.63}_{-0.62}$ & 1.221\nl
P7(W)   & 7.63$^{+0.25}_{-0.41}$ & 5.98$^{+0.22}_{-0.18}$ & 0.32$^{+0.06}_{-0.06}$ & 2.16$^{+0.76}_{-1.36}$ & 1.074\nl
P0(Center) & 5.76$^{+0.11}_{-0.11}$ & 4.89$^{+0.05}_{-0.05}$& 0.42$^{+0.02}_{-0.03}$ & 2.50$^{+0.00}_{-0.45}$ & 1.067\nl

\enddata
\tablenotetext{a}{Errors are 90\% confidence level}
\tablenotetext{b}{N$_{H}$ fixed at the corresponding Galactic value}
\end{deluxetable}
\clearpage

\begin{references}

\reference{}
Allen, S. W.,  Fabian, A. C., Johnstone, R. M., Nulsen, P. E. J., \& Edge, A.C. 1992, \mnras, 254, 51

                                \reference{}
Anders, E., \& Grevesse N. 1989, Geochimica et Cosmochimica Acta, 53, 197

				\reference{}
Arnaud, K. A., Mushotzky, R. F., Ezawa, H., Fukazawa, Y., Ohashi, T., 
    Bautz, M. W., Crewe, G. B., Gendreau, K. C., Yamashita, K., Kamata, Y., \&
    Akimoto, F. 1994, \apjl, 436, L67

                                \reference{}
Arnaud, K. A. 1996, in  Astronomical Data Analysis Software and Systems V,
    ASP Conf. Series volume 101, eds. Jacoby, G., \& Barnes, J., p.17

                                \reference{}
Bevington, P. R. 1969, Data Reduction and Error Analysis for the Physical
    Sciences (New York: McGraw-Hill), p.200
                           
			        \reference{}
Bird, C. M. 1993, PhD thesis, Minnesota University, Minneapolis

                                \reference{}
Branduardi-Raymont, G., Fabricant, D., Feigelson, E., Gorenstein, P., Grindlay, J., Soltan, A., \& Zamorani, G. 1981, \apj, 248, 55

                                \reference{}
Brunzendorf, J., \&  Meusinger, H. 1999, \aaps, 139,141

                            \reference{}
Dickey, J. M., \& Lockman, F. J. 1990, \araa 28, 215
  
                              \reference{}
Dottani, T., Yamashita, A., Ezuka, H., Takahashi, K., Crew, G., Mukai, K., \& the SIS 
team 1997, ASCA News 5, April. (see also heasarc.gsfc.nasa.gov/docs/asca/
gis\_sis\_effective\_area.html)

                                \reference{}
Dupke, R. A., \& Arnaud, K. A. 2000, \apj, submitted.

                \reference{}
Dupke, R. A., \& White, R. E. III 2000, \apj, 537, 123 

                \reference{}
Crone, M. M., Evrard, A. E., \& Richstone, D. O. 1996, ApJ, 467, 489


                \reference{}
Edge, A. C., Stewart, G. C., \& Fabian, A. C. 1992, \mnras, 258, 177

                \reference{}
Ettori, S., Fabian, A. C., \& White D.A. 1998, \mnras, 300, 837


                                \reference{}
Evrard, A. E. 1990, \apj, 363, 349;  

                                \reference{}
Evrard, A. E., Metzler, C. A., \& Navarro, J. F. 1996, \apj, 469, 494; 

				\reference{}
Eyles, C. J., Watt, M. P., Bertram, D., Church, M. J., Ponman, T. J., Skinner, G. K., Willmore, A. P., 1991, \apj, 376, 23

				\reference{}
Fabian, A. C.\& Daines, S. J. 1991, \mnras, 252, 17

                        \reference{}
Fitchett, M.J. 1988, in "Minnesota lectures on clusters of galaxies and large-scale structure", ed. Dickey, J. (San Francisco: Astronomical Society of the Pacific) p.143

                               \reference{}
Fritz, G., Davidsen, A., Meekins, J. F., \& Friedman, H. 1971, \apjl, 164, 81

                               \reference{}
Gomez, P. L., Loken, C., Roettiger, K., \& Burns, J. O. 2000, \apj, submitted

                              \reference{}
Hwang, U., Mushotzky, R. F., Burns, J. O., Fukazawa, Y., \& White, R. A. 1999, \apj, 516, 604

                              \reference{}
Idesawa, E., Asai, K., Ishisaki, Y., Kubo, H, Kubota, A., Makishima, K., Tamura, T., Tashiro, M., \& the GIS team 1995, heasarc.gsfc.nasa.gov/docs/asca/gain.html

                                \reference{}
Kaastra, J. S. 1992, An X-Ray Spectral Code for Optically Thin Plasmas, (Internal SRON-Leiden Report, updated version 2.0)

                                \reference{}
Katz, N., \& White, S. D. M. 1993, \apj, 412, 455;

                                \reference{}
Kowalski, M. P., Cruddace, R. G., Snyder, W. A., Fritz, G. G., Ulmer, M. P., \& Fenimore, E. E. 1993, \apj, 412, 489

                                \reference{}
Kowalski, M. P. 1994, in ``The Soft X-ray Cosmos'' ed. E. M. Schlegel \& R. Petre., 363

                                     \reference{}
Liedahl, D. A., Osterheld, A. L., \& Goldstein, W. H. 1995, \apjl, 438, L115
 
                                 \reference{}
McGlynn, T. A., \& Fabian, A. C. 1984, \mnras, 208, 709

                                   \reference{}
Markevitch, M., Ponman, T. J., Nulsen, P. E. J., Bautz, M. W., Burke, D. J.,David, L. P., Davis, D.,  Donnelly, R. H.,  Forman, W. R., Jones, C., Kaastra, J.,  Kellogg, E., Kim, D.-W.,  Kolodziejczak, J.,  Mazzotta, P.,  Pagliaro, A.,  Patel, S.,  VanSpeybroeck, L.,  Vikhlinin, A., Vrtilek, J., Wise, M., \&  Zhao P. 2000, \apj, submitted (astro-ph/0001269)

                                  \reference{}
Mewe, R., Gronenschild, E. H. B. M., \& Van den Oord, G.  H. J. 1985, \aaps, 62, 197

                                \reference{}
Mewe, R., Lemen, J. R., \& Van den Oord, G. H. J. 1986, \aaps, 65, 511

                                \reference{}
Mohr, J. J., Fabricant, D. G., \& Geller, M. J. 1993, \apj, 413, 492

                                \reference{}
Morrison, R., \& McCammon, D. 1983, \apj, 270, 119

                                \reference{}
Mushotzky, R. F., \& Szymkowiak, A. E. 1988, in Cooling Flows in Clusters and Galaxies, ed. A. C. Fabian (Dordrecht: Kluwer), 53

                                \reference{}
Navarro, J. F., Frenk, C. S., \& White, S. D. M. 1995 \mnras, 275, 720 

                                \reference{}
Pearce, F. R., Thomas, P. A., \& Couchman, H. M. P. 1994, \mnras, 268,953; 

                                \reference{}
Peres, C. B., Fabian, A. C., Edge, S. W., Johnstone, R. M., \& White, D. A. 1998, \mnras, 298, 416
 
                               \reference{}
Ponman, T. J., Bertram, D., Church, M. J., Eyles, C. J., Watt, M. P., Skinner, G. K., \& Willmore, A. P. 1990, \nat, 347, 450

                               \reference{}
Ricker, P. M. 1998, \apj, 496, 670

                                \reference{}
Roettiger, K., Burns, J. O., \& Loken, C. 1993, \apjl, 407, 53;

                                \reference{}
Roettiger, K., Burns, J. O., \& Loken, C. 1996, \apj, 473, 651

                                \reference{}
Roettiger, K., Loken, C., \& Burns, J. O. 1997, \apjs, 109, 307

                                \reference{}
Roettiger, K.\& Flores, R. 2000, \apj, 538, 92

                                \reference{}
Schindler, S., \& Muller, E. 1993, \aap, 272, 137 ;

                                \reference{}
Schwarz, R. A., Edge, A. C., Voges, W., Boehringer, H., Ebeling, H., \& Briel, U. G. 1992, \aap, 256, L11

                                \reference{}
Snyder, W. A., Kowalski, M. P., Cruddace, R. G., Fritz, G. G., Middleditch, J., Fenimore, E. E., Ulmer, M. P., \& Majewski, S. R. 1990, \apj, 365, 460

                                \reference{}
Takizawa, M., \& Mineshige, S. 1998, \apj, 499, 82; 

                               \reference{}
Takizawa, M. 1999, \apj, 520,514

                                \reference{}
Takizawa, M. 2000, \apj,  in press (astro-ph/9910441)

                               \reference{}
Tashiro, M., Fukazawa, Y., Idesawa, R., Ishisaki, Y., Kubo, H., Makishima, K., Ueda, Y., \& the GIS team 1995, ASCA News 3, August.

                                \reference{}
Tashiro, M., Kubota, A., Kubo, H., Ebisawa, K., \& the GIS team 1999, heasarc.gsfc.nasa.gov/docs/asca/gisnews.html (see also --- gain.html)

                               \reference{}
Ulmer, M. P., Cruddace, R. G., Fenimore, E. E., Fritz, G. G., \& Snyder, W. A. 1987,
\apj, 319, 118

                               	\reference{}
Ulmer, M.P., Wirth, G. D., \& Kowalski, M. P. 1992, \apj, 397, 430


                               \end{references}
                               
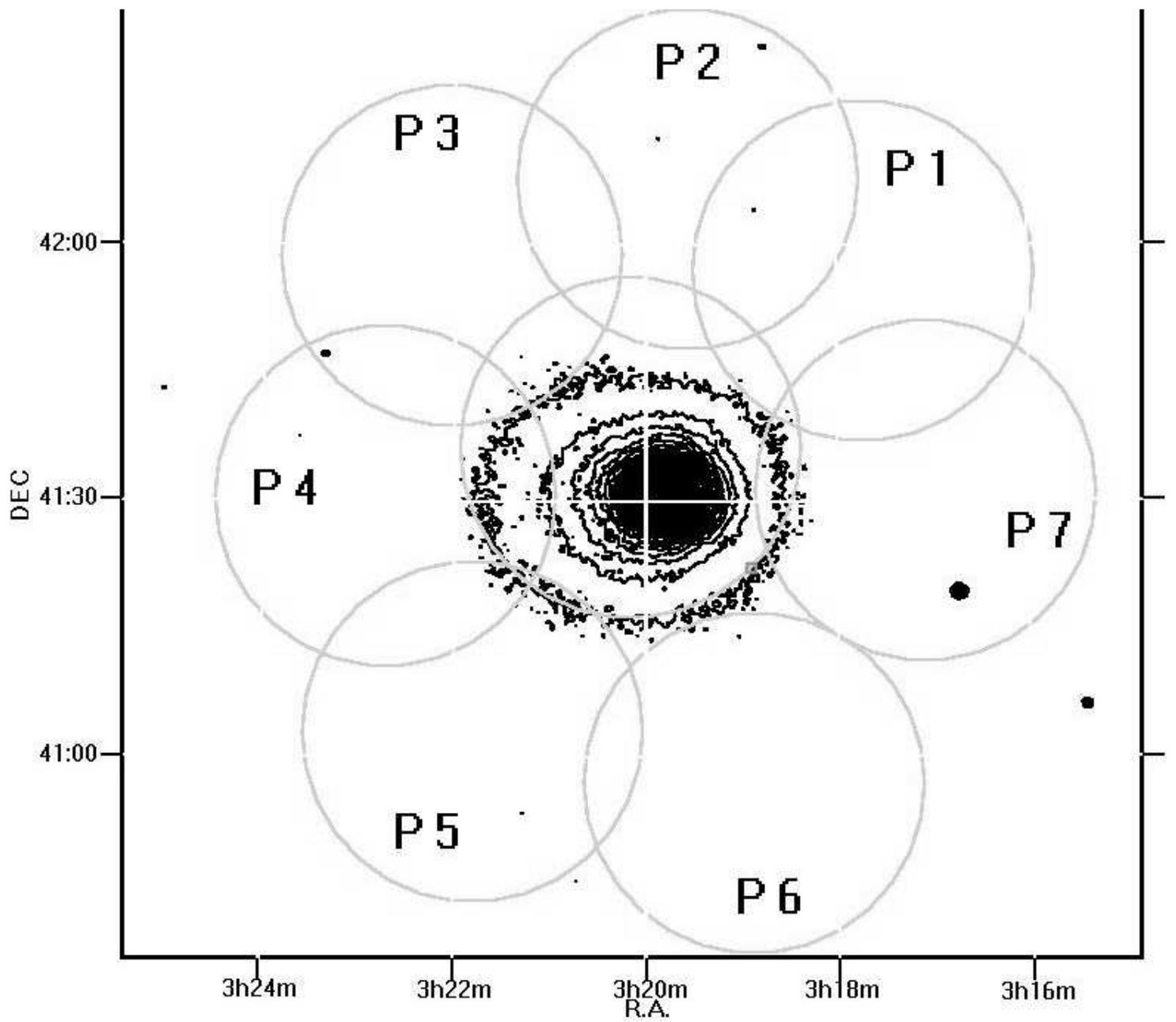
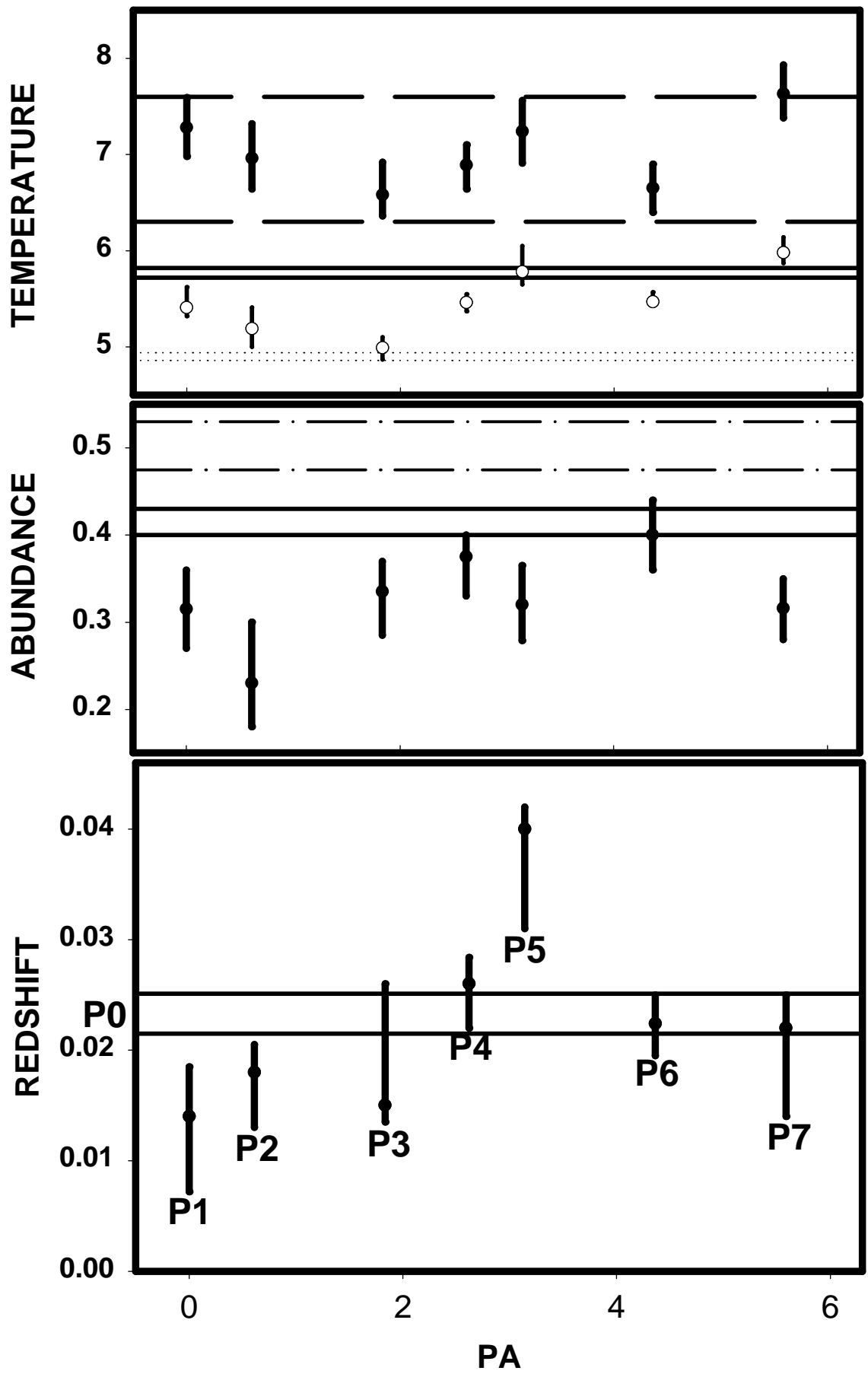
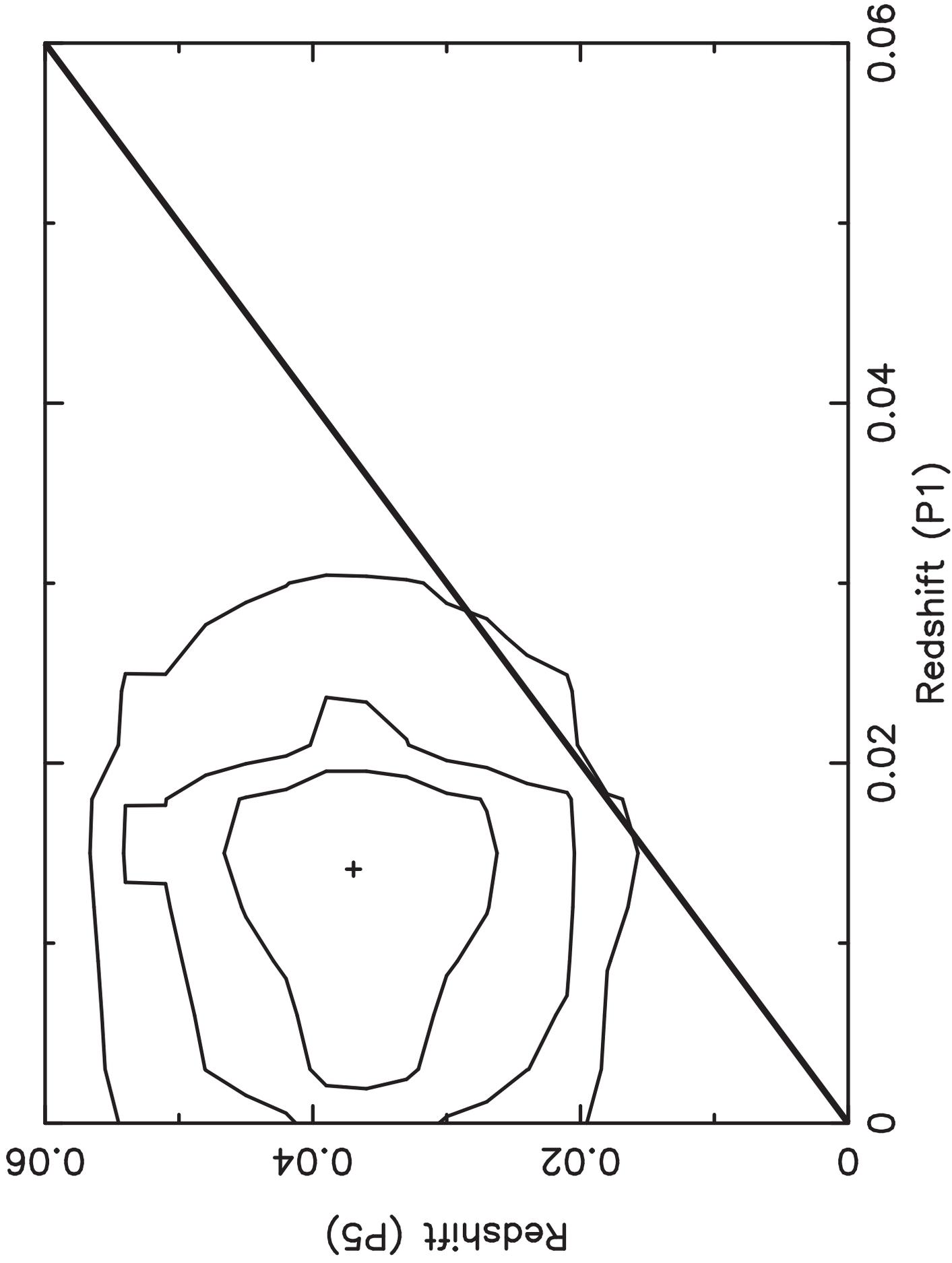
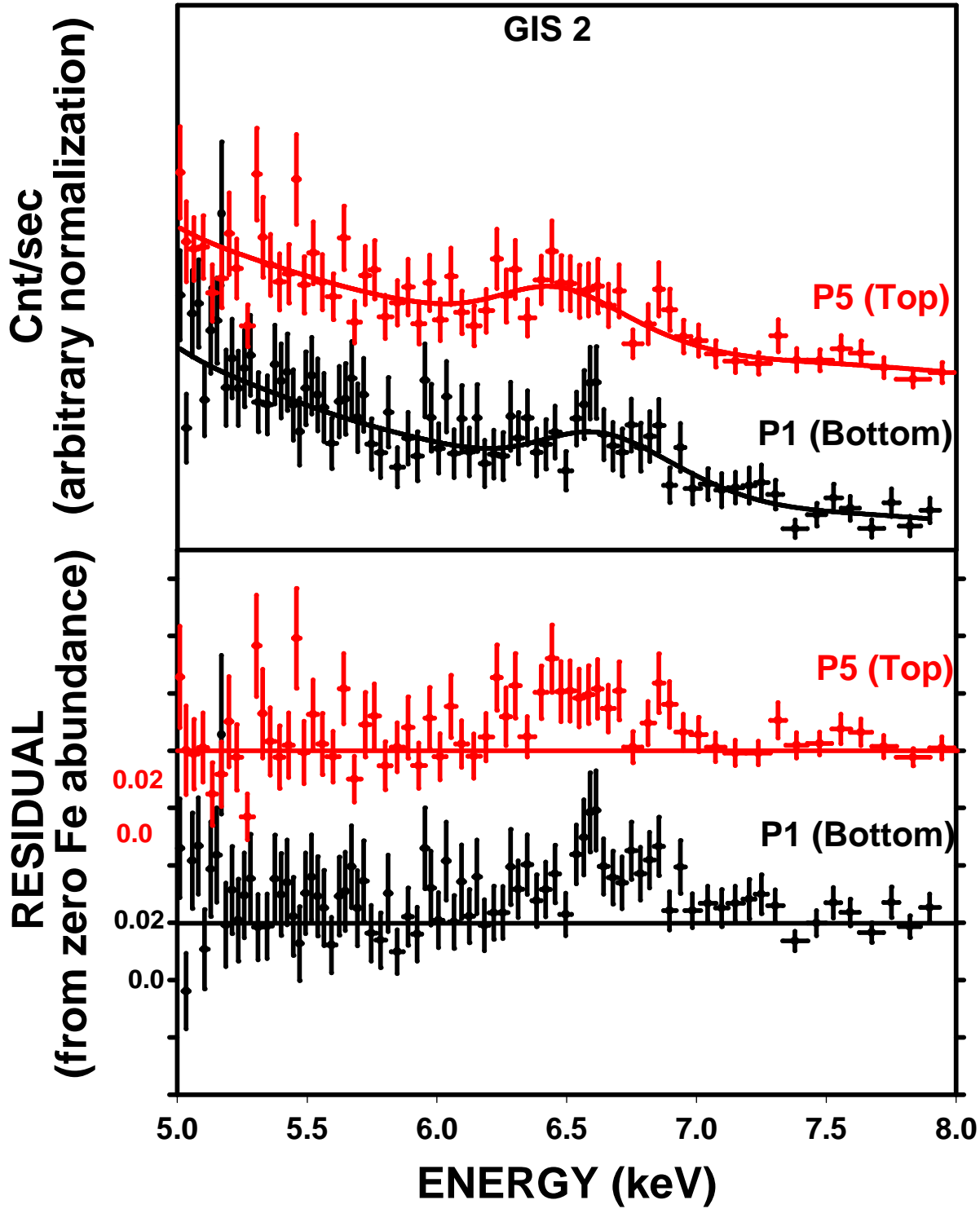
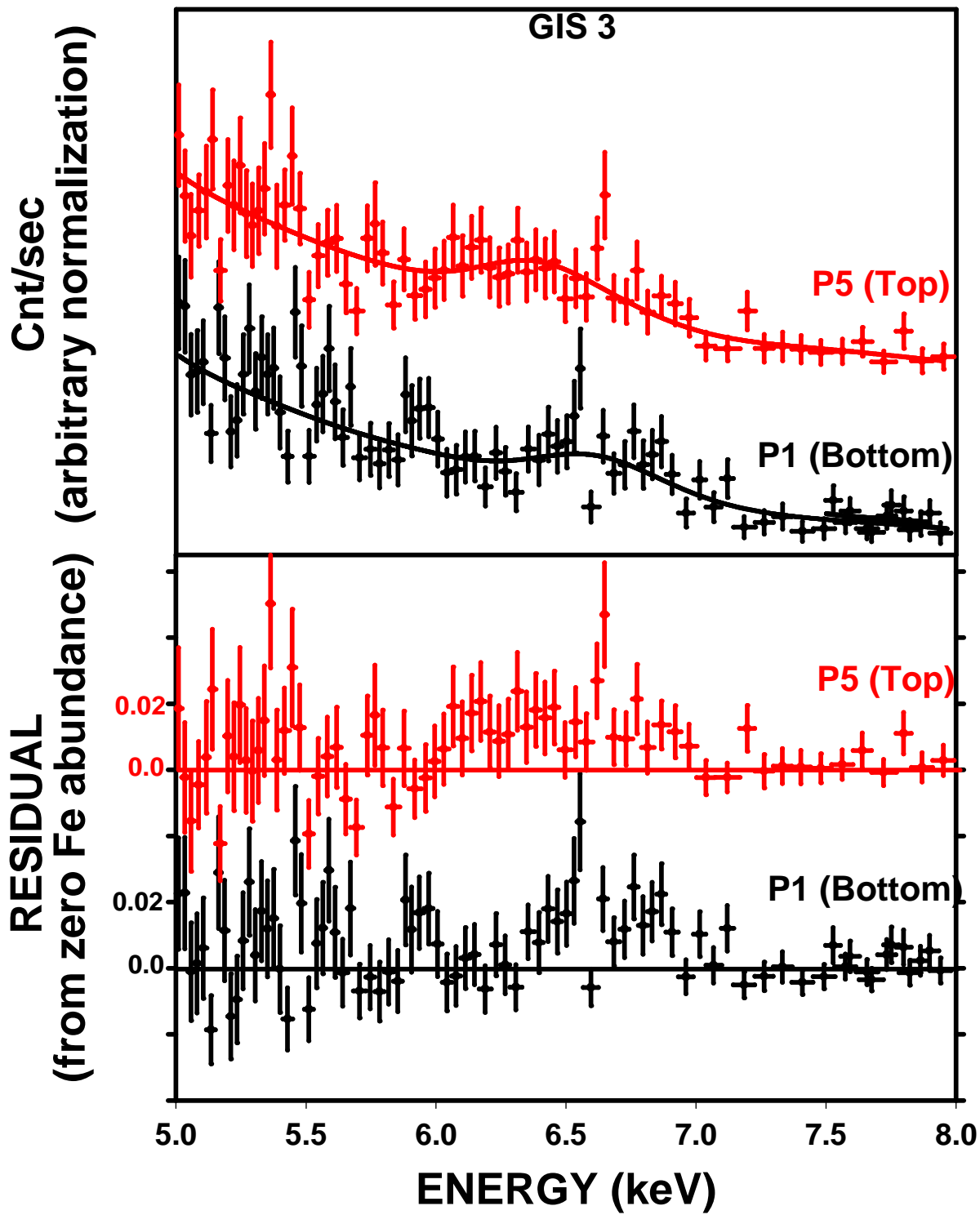
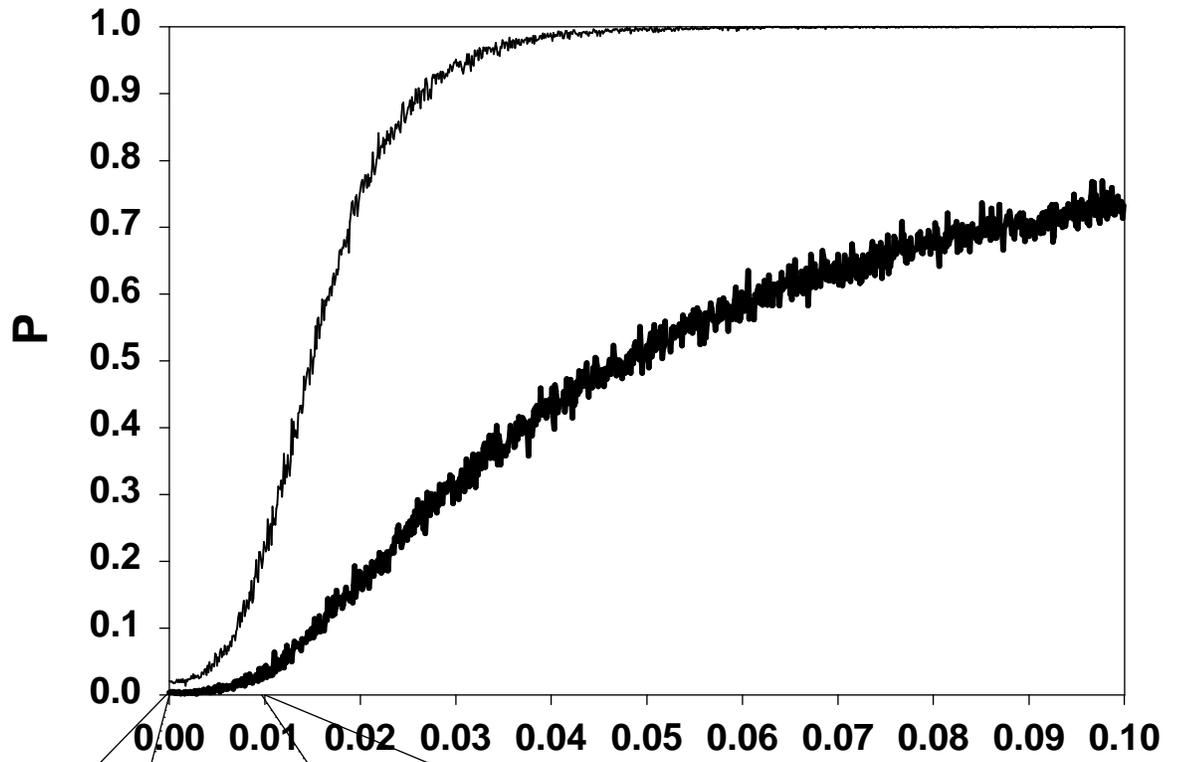
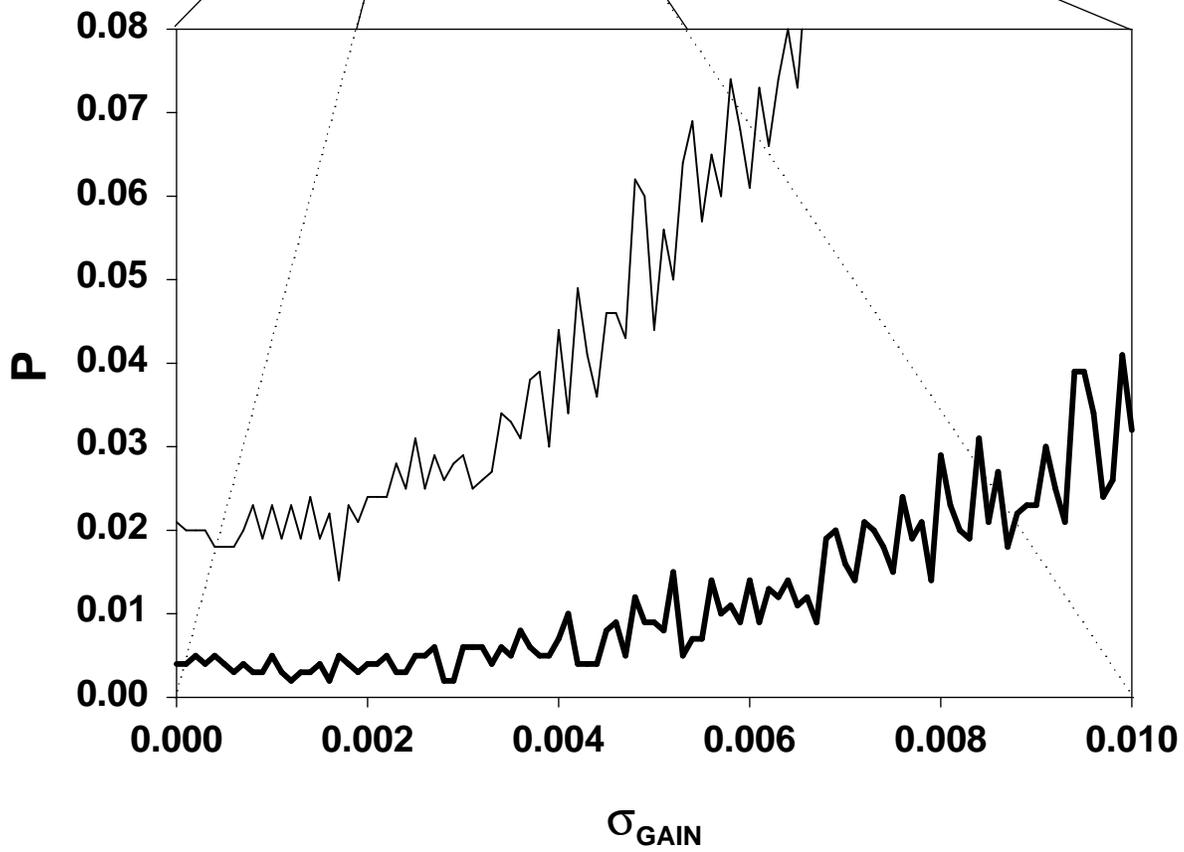
\end{document}